\documentclass[a4paper,11pt]{article}

\pdfoutput=1 

\usepackage{jinstpub} 

\usepackage{soul}
\newcommand{\khzcm}{\mathrm{kHz/cm^2}}
\newcommand{\hzcm}{\mathrm{Hz/cm^2}}

\title{\boldmath High rate, fast timing Glass RPC for the high $\eta$ CMS muon detectors}

\author[aa]{F.\,Lagarde, M.\,Gouzevitch, I. Laktineh, 
V.\,Buridon, X.\,Chen, C.\,Combaret, A.\,Eynard, L.\,Germani, G.\,Grenier, H.\,Mathez, L.\,Mirabito,  A.\,Petrukhin, A.\,Steen, W.\,Tromeur}
\affiliation[aa]{Institut de Physique Nucleaire de Lyon, Universite de Lyon, Universite Claude Bernard Lyon 1, CNRS-IN2P3, Villeurbanne, France}

\author[ab]{\\Y.\,Wang, A.\,Gong}
\affiliation[ab]{Tsinghua University, Beijing, China}

\author[ac]{, N.\,Moreau, C.\,de la Taille, F.\,Dulucq}
\affiliation[ac]{Omega-\'Ecole Polytechnique, Paris, France}

\author[b]{, A. Cimmino, S. Crucy, A.Fagot, M. Gul, A.A.O. Rios, M. Tytgat, N. Zaganidis}
\author[c]{, S. Aly, Y. Assran, A. Radi, A. Sayed}
\author[d]{, G. Singh}
\author[e]{, M. Abbrescia, G. Iaselli, M. Maggi, G. Pugliese, P. Verwilligen}
\author[f]{, W. Van Doninck}
\author[g]{, S. Colafranceschi, A. Sharma}
\author[h]{, L. Benussi, S. Bianco, D.Piccolo, F. Primavera}
\author[i]{, V. Bhatnagar, R. Kumari, A. Mehta, J. Singh}
\author[j]{, A. Ahmad, W. Ahmed, H., M. I. Asghar, I. M. Awan, R. Hoorani, S. Muhammad, H. Shahzad, M.A. Shah}
\author[k]{, S. W. Cho, S. Y. Choi, B. Hong, M. H. Kang, K. S. Lee, J. H. Lim, S. K. Park}
\author[l]{, M.S. Kim}
\author[m]{, S. Carpinteyro Bernardino, I. Pedraza, C. Uribe Estrada}
\author[n]{, S. Carrillo Moreno, F. Vazquez Valencia}
\author[o]{, L.M. Pant}
\author[p]{, S. Buontempo, N. Cavallo, M. Esposito, F. Fabozzi, G. Lanza, I. Orso, L. Lista, S. Meola, M. Merola, P. Paolucci, F. Thyssen}
\author[q]{, A. Braghieri, A. Magnani, P. Montagna, C. Riccardi, P. Salvini, I. Vai, P. Vitulo}
\author[r]{, Y. Ban, S.J. Qian}
\author[s]{, M. Choi}
\author[t]{, Y. Choi, J. Goh, D. Kim}
\author[u]{, A. Aleksandrov, R. Hadjiiska, P. Iaydjiev, M. Rodozov, S. Stoykova, G. Sultanov, M. Vutova}
\author[v]{, A. Dimitrov, L. Litov, B. Pavlov, P. Petkov}
\author[w]{, I. Bagaturia, D. Lomidze}
\author[x]{, C. Avila, A. Cabrera, J.C. Sanabria}
\author[y]{, I. Crotty}
\author[z]{, J. Vaitkus}

\affiliation[b]{Ghent university, Dept. of Physics and Astronomy, Proeftuinstraat 86, B-9000 Ghent, Belgium}
\affiliation[c]{Egyptian Network for High Energy Physics, Academy of Scientific Research and Technology, 101 Kasr El-Einy St. Cairo Egypt.}
\affiliation[d]{Chulalongkorn University, Department of Physics, Faculty of Science, Payathai Road, Phatumwan, Bangkok, THAILAND - 10330.}
\affiliation[e]{INFN, Sezione di Bari, Via Orabona 4, IT-70126 Bari, Italy.}
\affiliation[f]{Vrije Universiteit Brussel, Boulevard de la Plaine 2, 1050 Ixelles, Belgium.}
\affiliation[g]{Physics Department CERN, CH-1211 Geneva 23, Switzerland.}
\affiliation[h]{INFN, Laboratori Nazionali di Frascati (LNF), Via Enrico Fermi 40, IT-00044 Frascati, Italy.}
\affiliation[i]{Department of Physics, Panjab University, Chandigarh Mandir 160 014, India.}
\affiliation[j]{National Centre for Physics, Quaid-i-Azam University, Islamabad, Pakistan.}
\affiliation[k]{Korea University, Department of Physics, 145 Anam-ro, Seongbuk-gu, Seoul 02841, Republic of Korea.}
\affiliation[l]{Kyungpook National University, 80 Daehak-ro, Buk-gu, Daegu 41566, Republic of Korea.}
\affiliation[m]{Benemerita Universidad Autonoma de Puebla, Puebla, Mexico.}
\affiliation[n]{Universidad Iberoamericana, Mexico City, Mexico.}
\affiliation[o]{Nuclear Physics Division Bhabha Atomic Research Centre Mumbai 400 085, INDIA.}
\affiliation[p]{INFN, Sezione di Napoli, Complesso Univ. Monte S. Angelo, Via Cintia, IT-80126 Napoli, Italy.}
\affiliation[q]{INFN, Sezione di Pavia, Via Bassi 6, IT-Pavia, Italy.}
\affiliation[r]{School of Physics, Peking University, Beijing 100871, China.}
\affiliation[s]{University of Seoul, 163 Seoulsiripdae-ro, Dongdaemun-gu, Seoul, Republic of Korea.}
\affiliation[t]{Sungkyunkwan University, 2066 Seobu-ro, Jangan-gu, Suwon-si, Gyeonggi-do, Republic of Korea.}
\affiliation[u]{Bulgarian Academy of Sciences, Inst. for Nucl. Res. and Nucl. Energy, Tzarigradsko shaussee Boulevard 72, BG-1784 Sofia, Bulgaria.}
\affiliation[v]{Faculty of Physics, University of Sofia,5, James Bourchier Boulevard, BG-1164 Sofia, Bulgaria.}
\affiliation[w]{Tbilisi University, 1 Ilia Chavchavadze Ave, Tbilisi 0179, Georgia.}
\affiliation[x]{Universidad de Los Andes, Apartado Aereo 4976, Carrera 1E, no. 18A 10, CO-Bogota, Colombia.}
\affiliation[y]{Dept. of Physics, Wisconsin University, Madison, WI 53706, United States.}
\affiliation[z]{Vilnius University, Vilnius, Lithuania.}

\emailAdd{f.lagarde@ipnl.in2p3.fr}
\emailAdd{mgouzevi@cern.ch}
\emailAdd{laktineh@ipnl.in2p3.fr}

\abstract{The HL-LHC phase is designed to increase by an order of magnitude the amount of data to be collected by the LHC experiments.  To achieve this goal in a reasonable time scale the instantaneous luminosity would also increase by an order of magnitude up to $6 \cdot 10^{34}$\,cm$^{-2}$s$^{-1}$. 
 The region of the forward muon spectrometer
($|\eta| > 1.6$) is not equipped with RPC stations. The increase of the expected particles rate up to 2\,kHz/cm$^2$ ( including a safety factor 3 ) motivates the installation of RPC chambers to guarantee redundancy with the CSC chambers already present. The current CMS RPC technology cannot sustain the expected background level.
The new technology that will be chosen should have a high rate capability and provide a good spatial and timing resolution. A new generation of Glass-RPC (GRPC) using low-resistivity glass is proposed to equip at least the two most far away of the four high $\eta$ muon stations of CMS. First the design of small size prototypes and studies of their performance in high-rate particles flux are presented. Then the proposed designs for large size chambers and their fast-timing electronic readout are examined and preliminary results are provided.}


\arxivnumber{1606.01398} 

\collaboration[c]{\\ on behalf of CMS RPC collaboration}

\proceeding{XIII$^{\text{th}}$ Workshop on Resistive Plate Chambers and related detectors\\
  22-26 February, 2016\\
  Ghent University, Belgium}

\begin{document}
\maketitle
\flushbottom
\section{Introduction}
In the present CMS detector, all the muon stations are equipped with two kinds of muon detectors.  Drift Tubes (DT) and Resistive Plate Chambers (RPC) detectors are used to ensure a good redundancy in the barrel region.  In the endcap region, Cathode Strip Chambers (CSC) and RPC are used except in the stations of high eta region ($\eta > 1.6$) where only CSC detectors are present.
To guarantee a redundancy in this region and improve the muon trigger efficiency it is planned to add new chambers during long shut-downs LS2 and LS3. The projected increase of the LHC luminosity  up to $6 \cdot 10^{34}$\,cm$^{-2}$s$^{-1}$ during the HL-LHC phase suggests that new detectors with high rate capability are needed~\cite{upgrade}. 

\begin{figure}[!ht]
\begin{center}
    \includegraphics[width=0.60\textwidth]{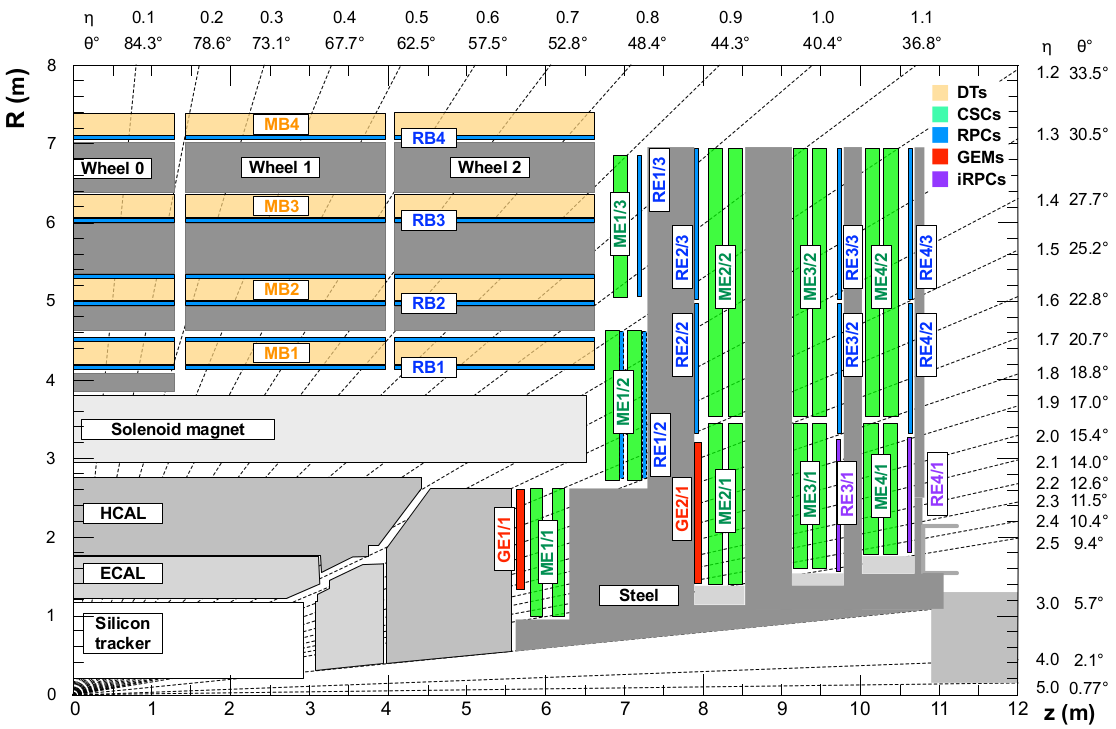}
  \end{center}
  \caption{Upgrade project of the CMS muon spectrometer with in red (purple) the two stations GE1/1 \& GE2/1 (RE3/1 \& RE4/1) proposed to be equiped by GEMs (RPCs).} \label{fig.upgrade}
\end{figure}

Figure \ref{fig.upgrade} summarizes the muon spectrometer upgrade project.
Gaseous electron multiplier (GEM) detectors are proposed to equip the two first (GE1/1 \& GE2/1) of these four high $\eta$ muon stations.  These high rate detectors provide the spatial resolution needed to solve the ambiguity that could affect the CSC when several particles are present. For the other stations (RE3/1 \& RE4/1), several RPC technologies are proposed. In this region the expected rate during HL-LHC program estimated with FLUKA~\cite{FLUKA1}\cite{FLUKA2} shall not exceed 2\,$\khzcm$ including a safety factor 3 \cite{upgrade}. Glass RPC is one of the proposed detectors. 

In its simple version a generic RPC detector is made of two resistive plates whose outer face is covered by a resistive coating. The two plates play the role of the electrodes.  The distance between the two electrodes is maintained constant using spacers.  A gas mixture circulates in between the two plates.  Applying high voltage (HV) on the two electrodes creates an electric field inside the gas gap. When a charged particle crosses the detector, it ionizes the gas molecules. Primary electron-ion pairs are produced and an avalanche is initiated under the electric field.  After the passage of the charged particle, the avalanche charges are absorbed through the resistive electrodes. The higher the resistivity of the electrodes the longer the time needed to evacuate the charge. 

During this process the electric field is locally diminished and passage of other charged particles may go undetected.  Once the charges are absorbed, the local electric field is then restored to its initial value. The resistivity of the electrodes used to eliminate possible sparks is also responsible of the limited RPC particle rate detection capability. 
 
A solution to overcome the GRPC detectors limited particle rate capabilities is to use low resistivity glass plates as electrodes. The new glass, developed by Tsinghua University, has a resistivity of the order of $10^{10}\,\Omega\cdot$cm and a very high surface uniformity, with a roughness below 10\,nm~\cite{TGRPC}\cite{Wang:2010bg}.  Low resistivity is an important ingredient to reach high rate capability since it allows the avalanche charges produced by the passage of a charged particle to be absorbed more quickly through the electrodes. The absorption can also be made faster by reducing the thickness of the electrodes. Thin but stiff glass plates represent then a good option. Another important element in reaching high rate capabilities is to reduce the  charge produced by the avalanche. This  is possible by reducing the gas gap separating the two electrodes. It is however compulsory to equip the GRPC in this case with low noise electronic readout.  

Two kinds of GRPC detectors are proposed. The first is made of two single-gap GRPC separated by an electronic board hosting the pick-up strips. The second that aims to achieve excellent time resolution will be made in the same way but  with  multigap GRPC similar to the ones described in~\cite{Wang:2010bg}. The number of gaps in the second scenario will be determined by the time resolution one would like to achieve to reduce the noise and to fulfill requirements from physics study.

In this paper we will present the results obtained in beam tests at CERN PS, SPS anf GIF++ with few small single-gap GRPCs made with Tsinghua Low Resistivity (LR) glass. They are equipped with either pick-up strips or pads. Section 2 presents the GRPC structure and the GRPC performance at CERN PS, SPS and GIF++, it is a detailed extension of an earlier proceeding pulished in Ref.~\cite{VCI2016}. In section 3, the performance of large two single-gap GRPC detectors are studied with cosmic rays. Finally, section 4, gives a short description of the electronic readout proposed to achieve high precision time resolution as well as the preliminary results obtained with such system.  
\section{ Small GRPC chambers }
\subsection{ GRPC structure}
\begin{figure}[htb]
\begin{center}
\includegraphics[width=0.70\textwidth]{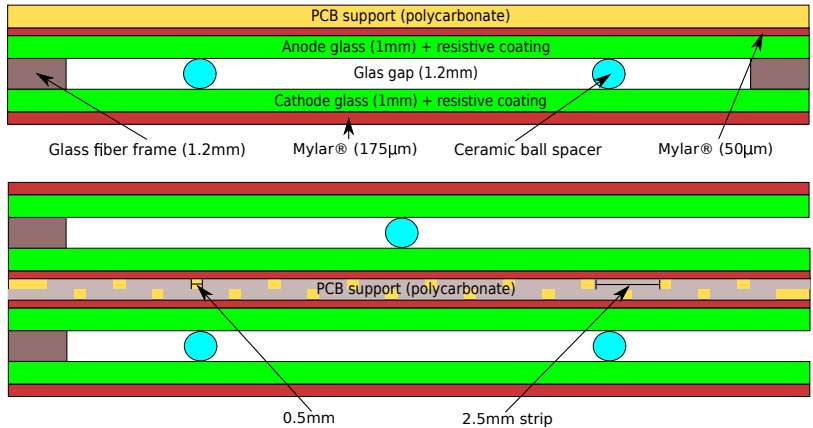}
\caption{Scheme of a small single gap GRPC chamber (top) and double-gap chamber (bottom).\label{fig.scheme}}
\end{center}
\end{figure}

The dimensions of the first prototypes are constrained by the largest size of the low resistivity glass plates that could be produced currently: $30 \times 32\,{\rm cm^{2}}$~\cite{TGRPC}. Few plates were used to build small prototypes sketched in Fig.~\ref{fig.scheme} (top): a gas gap of $1.2$\,mm separates two $1$\,mm thick low resistivity glass plates covered with a colloidal graphite coating (surface resistivity of about a few M$\Omega/\Box$). Spacers made of glass fiber and ceramic are used to maintain uniform the distance between the two plates. To operate the detectors a gas mixture from the SDHCAL project made of TFE(93\%), CO2(5\%) and  SF6(2\%) is used~\cite{Prototype}. Except for the glass electrode, the GRPCs are identical to those described in Ref.~\cite{small}.

The signal was collected using two kinds of PCB electronic plates. The first is equipped with $1\times1$\,cm$^2$ pads read by 24 64-channels HARDROC ASICs~\cite{hardroc}. 
The second is equipped on each side with 128 parallel strips.
The strips have the same direction on each side, are read by 4 ASICs, have a pitch of $d=2.5$\,mm, are separated by $0.5$\,mm and have an impedance of 24\,$\Omega$. The strips of one side are shifted by 1\,mm with respect to those of the other side in the direction perpendicular to the strips one (see  Fig.\ref{fig.scheme} (bottom)). This configuration, referred to as double-gap, is designed to increase the spatial resolution by looking at the coincidence of fired strips in the two layers. 

\subsection{Small GRPC performance in test beams}
\begin{figure}[htp]
\begin{center}
\includegraphics[width=0.46\textwidth]{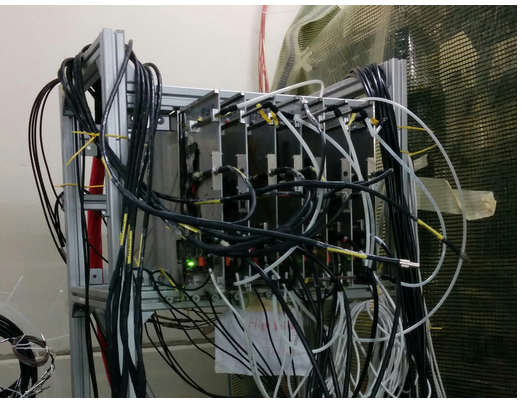}
\includegraphics[width=0.50\textwidth]{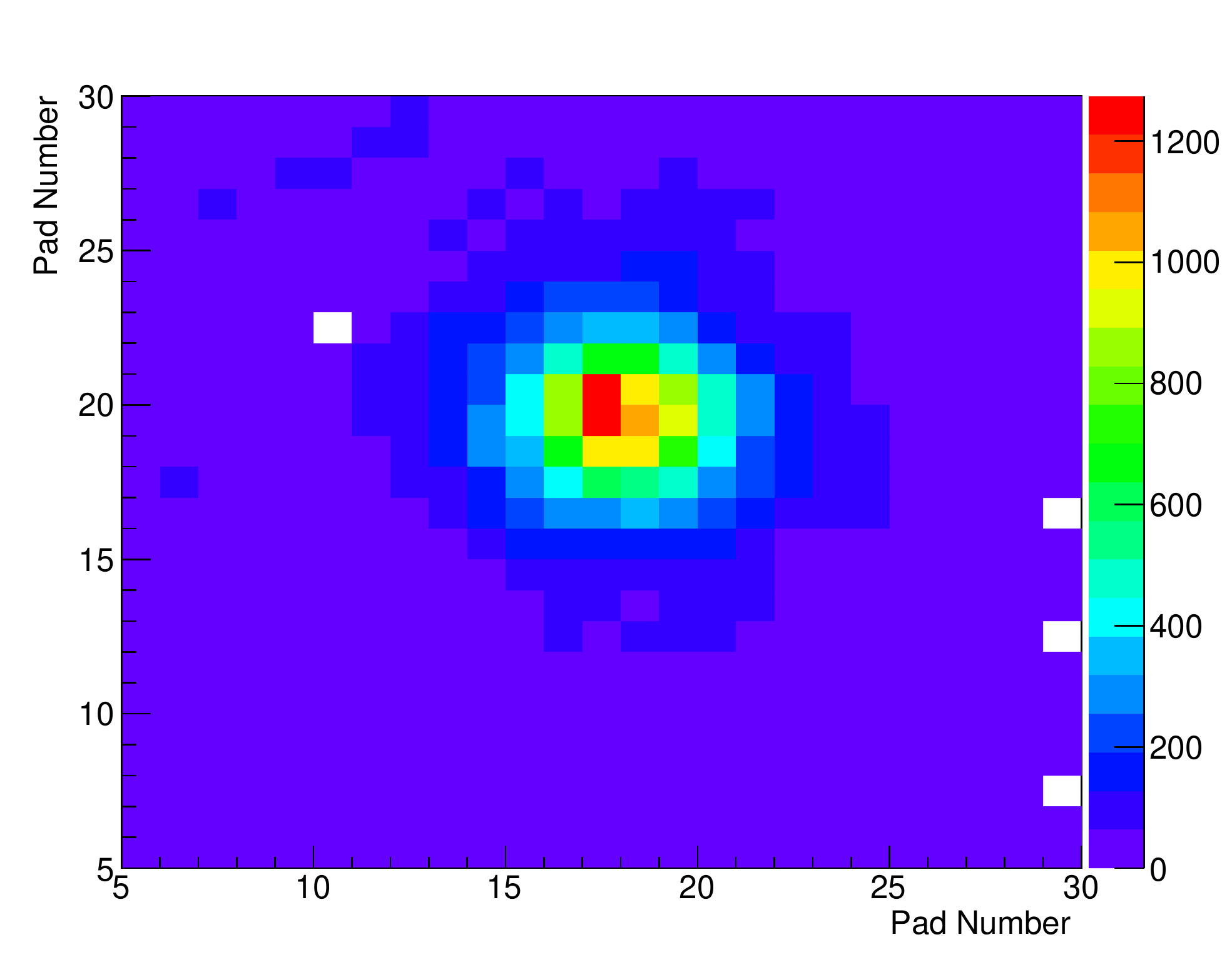}
\caption{Picture of the GRPC telescope (left) and an example of the beam profile obtained with a pad readout at $\approx 3\khzcm$ seen in a Tsinghua LR GRPC chamber (right). The $z$-axis is in number of particles measured per second. The profile is not corrected for efficiency.}
\label{fig.telescope}
\end{center}
\end{figure}

The performance of a small GRPC detector was validated for the first time in an electron beam at DESY~\cite{HRate}. More recently in 2014 and 2015, the chambers were exposed to a more intensive, wide and energetic $\mu/\pi$ beams in the CERN-PS and CERN-SPS lines. 
A telescope was built from several small chambers, four of which are Tsinghua LR GRPCs (see Fig.~\ref{fig.telescope} (left)). Additional small chambers made with float glass were also added to compare the behavior of the two kinds of GRPC in high rate conditions.  For the two beam tests a set of scintillators readout by photo-multipliers (PMTs) were used to measure the rate. 

The GRPC efficiency are measured in two ways. The first method uses the information provided by the coincidence signal of the different PMTs to tag the incoming particles and then the information recorded in the ASICs of the studied GRPC to check the presence of fired pads or strips.
In a second approach, the different fired pads (or strips) in the detectors other than the one in the chmaber under study are recorded according to their time of arrival.  Those whose time arrival is within 3 time slots of 200\,ns (the ASIC clock period) are gathered. If their number is higher than 3, then a $\chi^2$ test is performed to see if their positions are compatible with a straight line track. Fired pads or strips are searched within 3.0\,cm around the estimated impact point of the track in the studied detector. If some are found the detector is considered to be efficient and their number is recorded as an estimator of the cluster size.
The efficiency of the two methods are compared and found to be compatible. The final results are provided with the track method because of a higher space resolution associated to this approach. In Fig. \ref{fig.telescope} (right) we profit from the telescope resolution to build a beam profile with a pad readout. The $x$- and $y$-axes are given in units of pads that are $\approx 1.0$\,cm wide. The beam has a Gaussian shape with widths of 1.4~cm along $x$ and 1.6~cm along $y$. The SPS beam illuminates a rather narrow region of $6 \times 6\,{\rm cm^{2}}$ which contains 90\% of the incoming particles.

\begin{figure}[h]
\begin{center}
\includegraphics[width=0.63\textwidth]{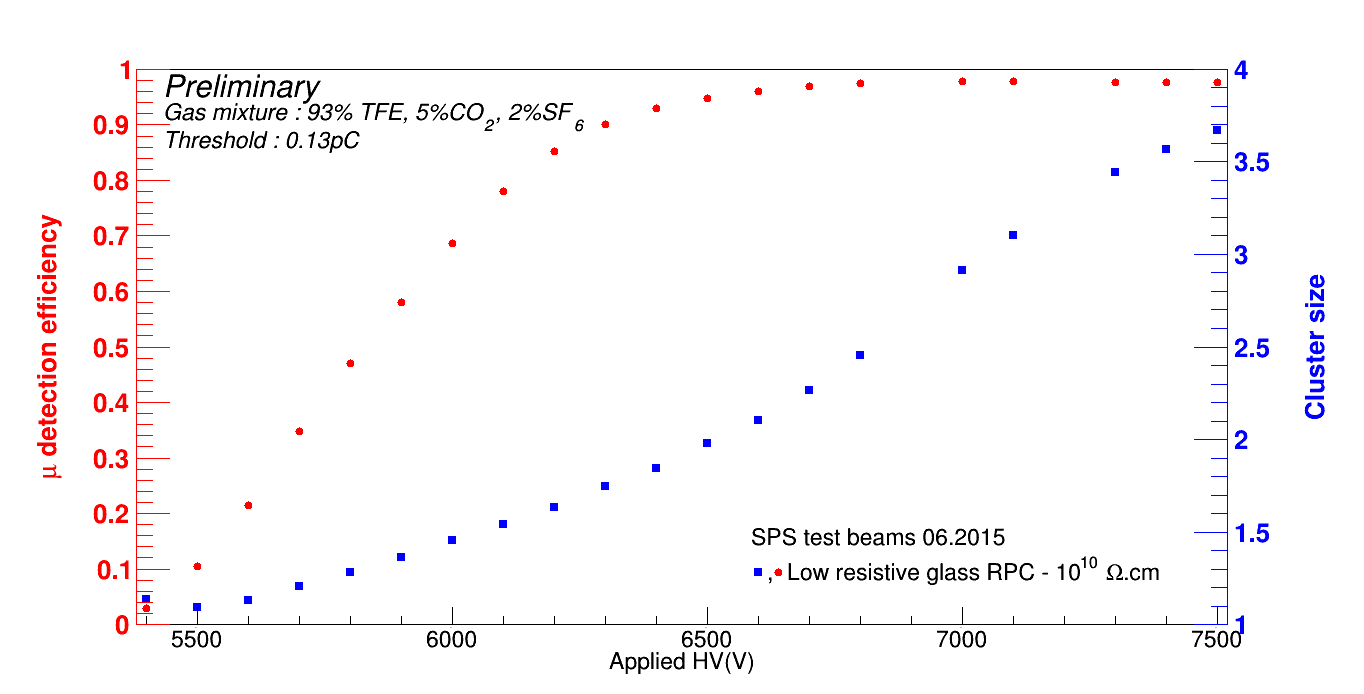}
\hspace*{-0.3cm}
\includegraphics[width=0.34\textwidth]{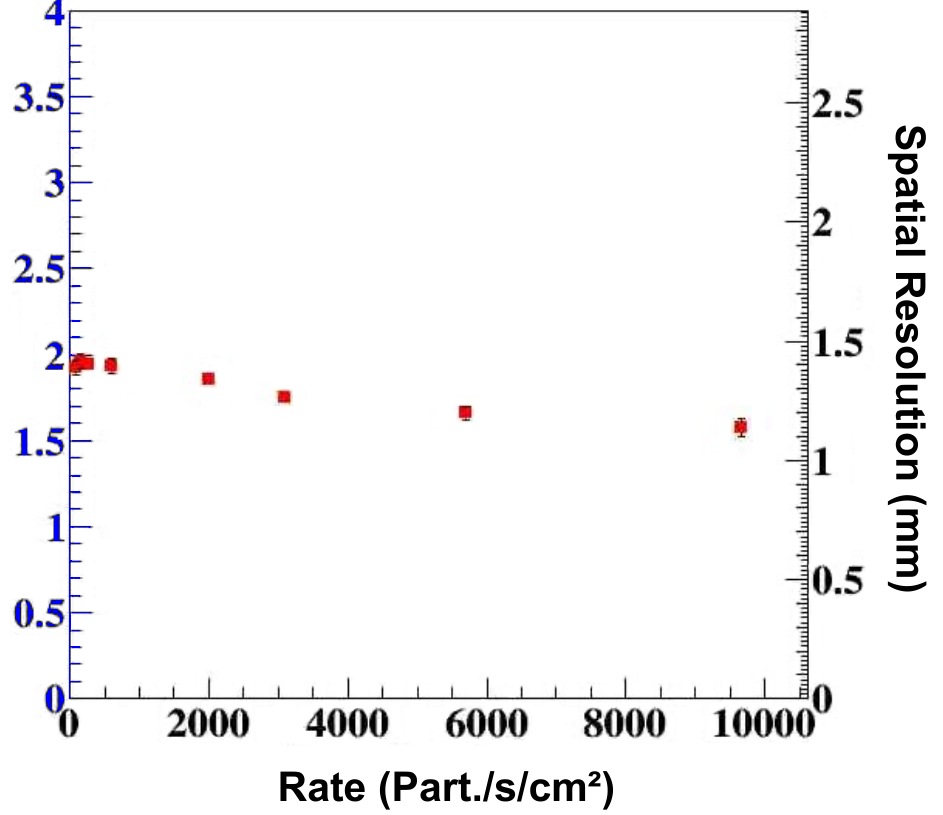}
\caption{Efficiency and cluster size of one single gap detector as a function of the applied HV measured in SPS using a pad readout (left). Spatial resolution of double-gap GRPC (using strip readout) as a function of the mean particles rate (right).}
\label{efficiency}
\end{center}
\end{figure}

In Fig.~\ref{efficiency} (left)  the evolution of the average efficiency and cluster size of one of the single gap detectors with pad readout is shown as a function of the applied HV during the SPS beam test. An efficiency plateau is reached around 6.7\,kV with a relatively small cluster size of 2.2 hits. The noise rate of the chambers and the electronics has been estimated by quantifying the rate of hits not associated to a track and found to not exceed $2.0\,\hzcm$ for both Tsinghua LR and float GRPC chambers.

The cluster size is used to estimate the spatial resolution of chambers equipped with a pick-up strips readout.  
Since the readout is purely digital, in case only strips of one layer are fired the spatial resolution is estimated as $ \sigma = N d  /\sqrt{12}$, where $N$ is the number of fired strips and  $d$ is the pitch. When the fired strips belong to both layers, only overlapped strips are considered and we can use a more precise estimator of the resolution, $\sigma = D/\sqrt{12}$,  where $D$ is  the distance between the two far edges of the overlapped strips.  
The spatial resolution as function of the rate is shown on Fig.~\ref{efficiency} (right) for an operating voltage of 6.9~kV. 
The resolution improves from $1.5$\,mm at low rate down to 1.1\,mm at 10\,$\khzcm$. This trend can be explained by a significant surface charge accumulation at high rate that results in less gain, that in turn effectively reduces the cluster size. 
The mean particle rate was estimated by dividing the total number of incoming particles per second by th $6\times 6\,{\rm cm^{2}}$ area used for the measurements.
Other methods can be used to estimate the mean rate~\cite{Wang:2013gga}. 
Our rate estimate $\phi$ is related to the rate estimate $\bar{\phi}$ using the method of~\cite{Wang:2013gga} by the relation $\phi=0.7 \bar{\phi}$.

\begin{figure}
\begin{center}
\includegraphics[width=0.95\textwidth]{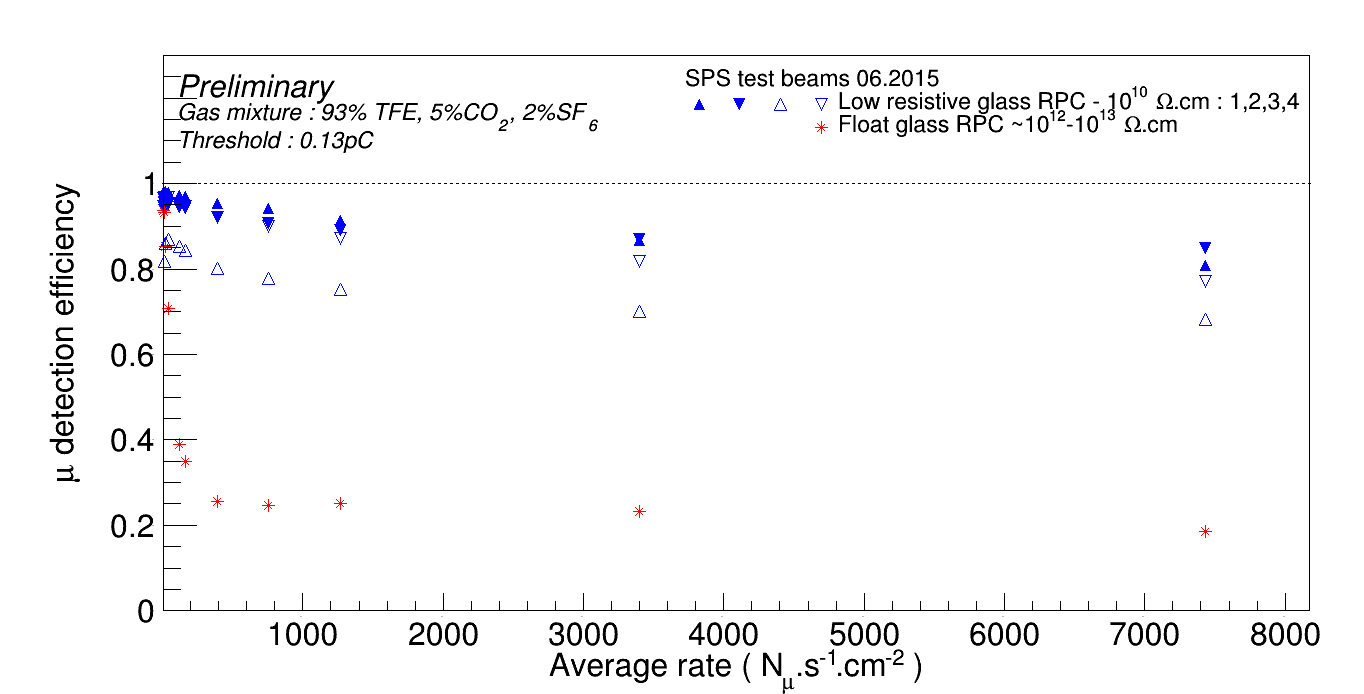}
\caption{Efficiency of the five detectors with a pad readout as a function of the mean particle rate in SPS.}
\label{fig:5Chamberseff}
\end{center}
\end{figure}

\begin{figure}[h]
\begin{center}
\includegraphics[width=0.95\textwidth]{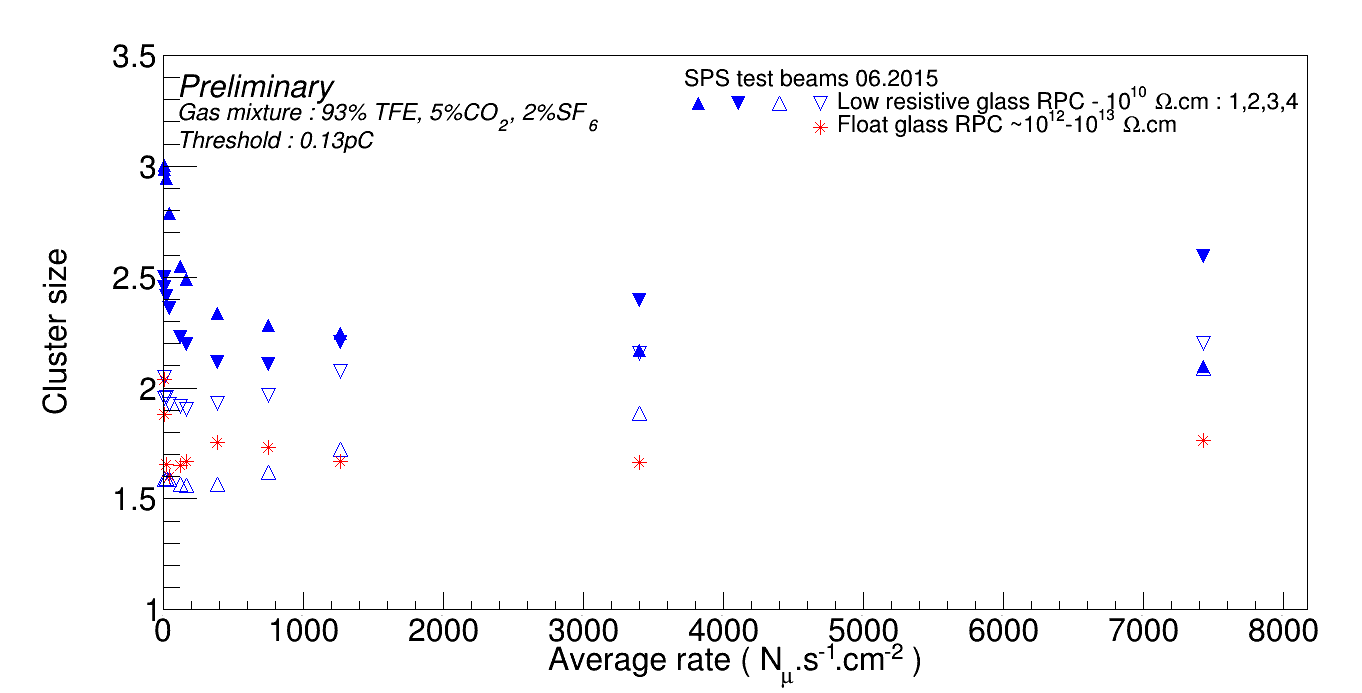}
\caption{Cluster size of the five detectors with a pad readout as a function of the mean particle rate in SPS.}
\label{fig:5Chambersmult}
\end{center}
\end{figure}

The efficiency and the cluster size of the the five detectors with a pad readout as a function of the particle rate are shown in Figs.~\ref{fig:5Chamberseff} and \ref{fig:5Chambersmult}. We observe that at low particle rate all the five chambers are very efficient ( $\epsilon > 90\%$), but the efficiency of the float glass RPC drops dramatically down to $10\%$ when the rate exceeds $0.1\,\khzcm$ while the four Tsinghua LR GRPC chambers keep being efficient at high rate albeit a small efficiency drop.
One of the low-resistivity GRPC exhibits a smaller efficiency than the three others. We track this inefficiency to the presence of dead channels in the electronic readout that was not corrected when estimating the efficiency. Nevertheless this chamber shows an identical efficiency trend as function of the particles rate up to a normalization factor close to unity. 

\subsection{Small GRPC performance in GIF++} 
Starting from July 2015 the telescope was installed in the GIF++ facility located in H4 line of SPS~\cite{GIFPP}. This facility is designed to emulate the harsh background conditions of the HL-LHC collisions for the muon detectors. A source of $10$\,TBq of $^{137}$Cs irradiates the chambers with $\gamma$ rays of 662\,keV. The advantage of this setup with respect to the SPS-only test beam is a more uniform immersion of the chamber into the background. The telescope is irradiated during long periods (many months) to collect a large cumulated charge. 

In between those periods a muon beam of several thousands of $\mu$\,/\,spill is used to test the efficiency of the chambers and monitor the aging process. The source is supplied with a system of movable lead attenuators that allows a reduction of the rate by factors between 1 and $10^{-5}$ in several steps, in particular during the test beam periods.

\begin{figure}[htp]
  \begin{center}
    \includegraphics[width=0.85\textwidth]{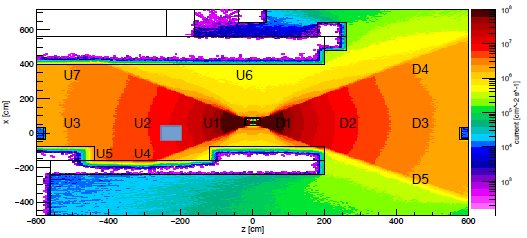}
  \caption{The irradiation map of the GIF++ facility. The blue square between the points U1, U2 and U4 indicates the approximative position of the telescope. \label{fig.position}}
  \end{center}
\end{figure}

The position of the GRPC telescope in GIF++ is sketched on Fig.~\ref{fig.position}. In this region the expected photon rate without attenuators is estimated to be of the order of $1.5\times 10^7.\gamma$.s$^{-1}$.cm$^{-2}$~\cite{GIFPP_RATE}. 
Due to the distance, this rate is reduced by few \% for the telescope's chambers further from the source. It can also vary by few \% when other setup, located between the telescope and the source are screening the later.

\begin{figure}[!htp]
\begin{center}
\includegraphics[width=0.85\textwidth]{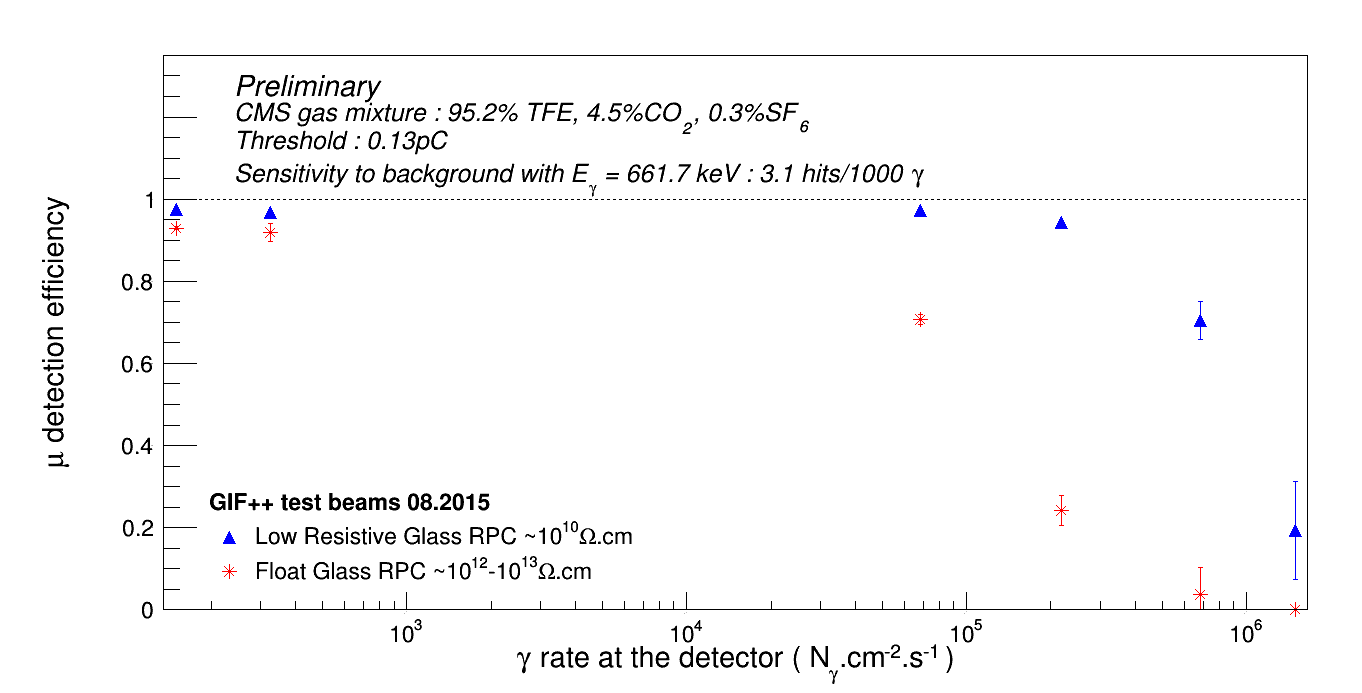}
\caption{The efficiency of a float and Tsinghua LR glass chamber as function of the $\gamma$ rate obtained using the attenuators.}
\label{fig::GIFPP}
\end{center}
\end{figure}

In August 2015, before the beginning of the first aging period, a test beam data sample was collected to estimate the initial efficiency of the chambers, shown in Fig.~\ref{fig::GIFPP}. Data samples were collected with source attenuator factors ranging from $3.3$ to $46000$ and GRPC HV set to 7~kV. 

The probability, $r_c = 0.31\%$, for a $\gamma$ to initiate a cluster in float glass RPC, is obtained with a GEANT4~\cite{GEANT4} based simulation of the telescope. Using this conversion factor, the maximum gamma rate translates to a GRPC hit rate of the order of 40\,$\khzcm$. Using $r_c$ to convert the $\gamma$ rate to a GRPC induced noise rate, Fig.~\ref{fig::GIFPP} shows that one can estimate that the single-gap float GRPC efficiency declines for induced background rate above 0.6\,$\khzcm$ and drops to 0 around 2\,$\khzcm$. 
Meanwhile, if we assume the same conversion rate for the Tsinghua LR GRPC, its efficiency is above 90\% for induced noise rate of 0.6\,$\khzcm$, is 70\% at 2\,$\khzcm$ and decreases down to 20\% for 6\,$\khzcm$.
This is significantly lower than the sustained rate observed at the SPS (see Fig.~\ref{fig:5Chamberseff}), suggesting the $r_c$ conversion factor might be higher for Tsinghua LR GRPC. 
An other possible explanation of this difference is the proportion of SF6.
In SPS, the gas mixture from SDHCAL design for the ILC project with 2\% of SF6 was used~\cite{Prototype}, while in GIF++ the standard CMS mixture was used with 7 times less SF6. This latter gas, known for is greenhouse effects, is used as electron quencher. A lower fraction of SF6 leads to a larger charge produced during the showers and, in turn, a stronger screening effect inside the detector reducing its efficiency.

These results show that the single gap Tsinghua LR GRPCs by themselves already fulfil the CMS HL-LHC upgrade requirements: if the single gap efficiency is $\epsilon \approx 70\%$, then the double-gap efficiency, $\epsilon_2 = 1- (1-\epsilon)^2 > 90\%$.
\section{ Large GRPC chambers}
The small chambers are important to validate the properties of the Tsinghua LR GRPC but their size is much smaller than the size of CMS chambers. Therefore it is important to produce a large trapezoidal prototype for high $\eta$ stations and test it. The limitation in size of the low resistivity glass plates requires to find an efficient and robust way to assemble the small plates. Two different assembling methods have been tried.

\begin{figure}[htp]
\begin{center}
\includegraphics[width=0.42\textwidth]{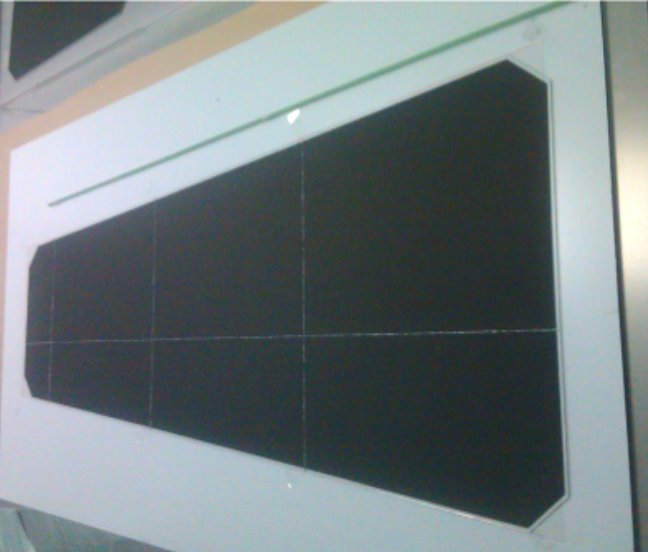}
\includegraphics[width=0.478\textwidth]{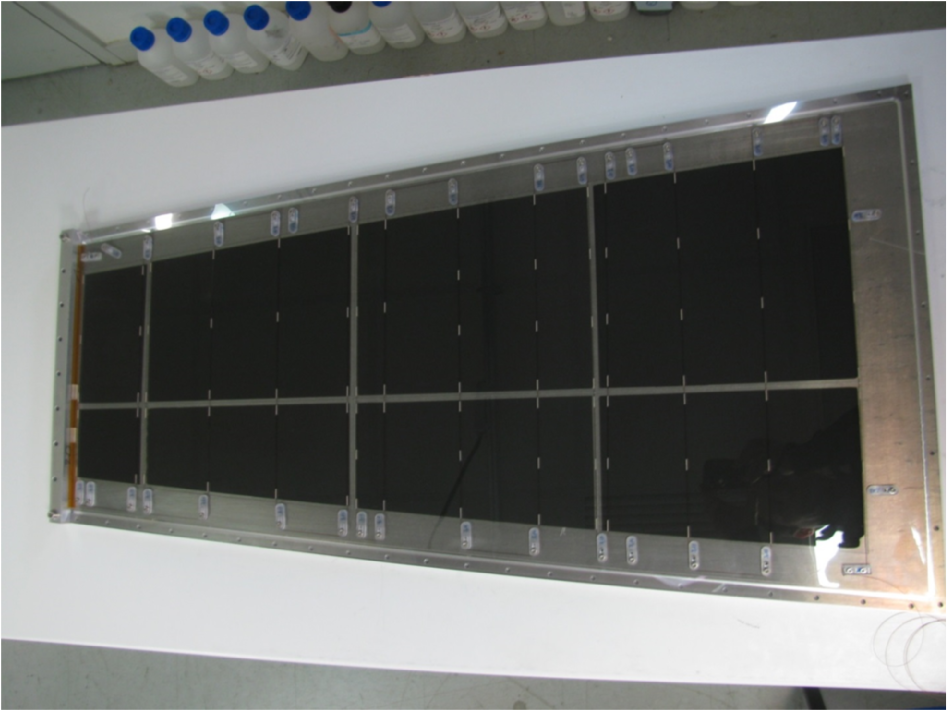}
\caption{Large size trapezoidal GRPC chamber built using glue (left) and mechanical fixation only (right).}
\label{Large_photo}
\end{center}
\end{figure}

The first one consists of  assembling the glass plates of the same thickness by RHODORSIL\textregistered \, CAF-4 glue before applying the coating (Figure \ref{Large_photo} (left)).  This allows to build gas-tight single gap detectors of large size similar to the small ones with limited dead zones. Using this method,  a detector including two single-gap GRPC is built and cast into an aluminum cassette. A PCB  with a $\approx 1.0$\,cm pitch is inserted between the two single gap GRPC. The strips are connected using coaxial cables to a test board hosting one HARDROC ASIC. 

The second method, called mechanical fixation, is proposed to avoid the usage of glue that could suffer from hard radiation (Figure \ref{Large_photo} (right)). The glass tiles are maintained mechanically by the holding aluminium cassette. Very thin copper tape are used to electrically connect the small glass together, forming an electrode. The gas gap between the two electrode is obtained using fishing lines. Two large GRPC are built in this way and as before a PCB with pick-up strips is inserted in between. In this design, the gas tightness is done in the cassette rather than in the space between the electrodes. A PMMA plate with the same size as the detectors is equipped with few springs and put in contact with the detectors in the cassette. The resulting compression effect ensures the plating of the glass tiles on the PCB. 

\begin{figure}[htp]
\begin{center}
\includegraphics[width=0.80\textwidth]{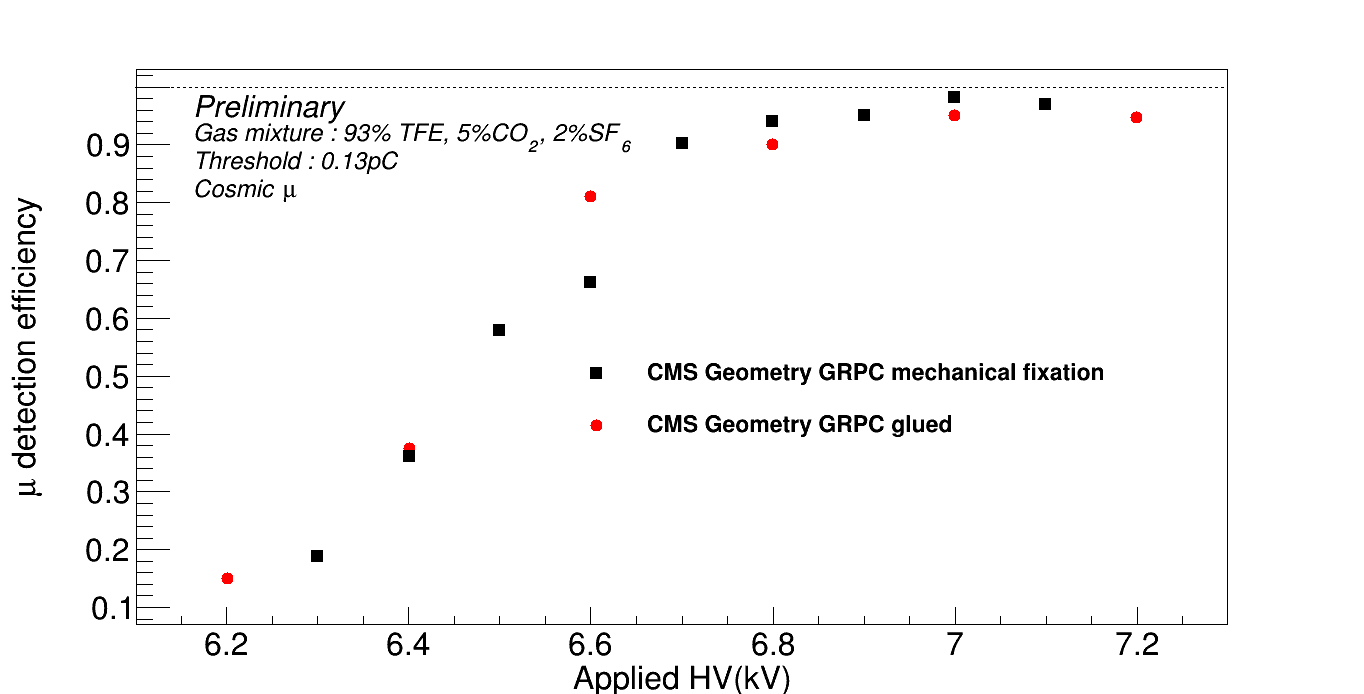}
\caption{Efficiency of the large detectors as a function of the applied HV.}
\label{fig::Large_eff}
\end{center}
\end{figure}

Both cassettes were placed in a cosmic test bench to measure their efficiency. Fig. \ref{fig::Large_eff} shows the efficiency of the large detector as a function of the applied HV. An efficiency plateau around 96\% is reached at 7\,kV for both chambers. The current-HV dependance exhibited a similar behavior for both chambers. The mechanical design features a smaller leakage current with respect to the glued one. This is because of the absence of the frame that ensures the gas tightness in that case. 

\section{Fast timing electronics}
\begin{figure}[htp]
\begin{center}
\includegraphics[width=0.48\textwidth]{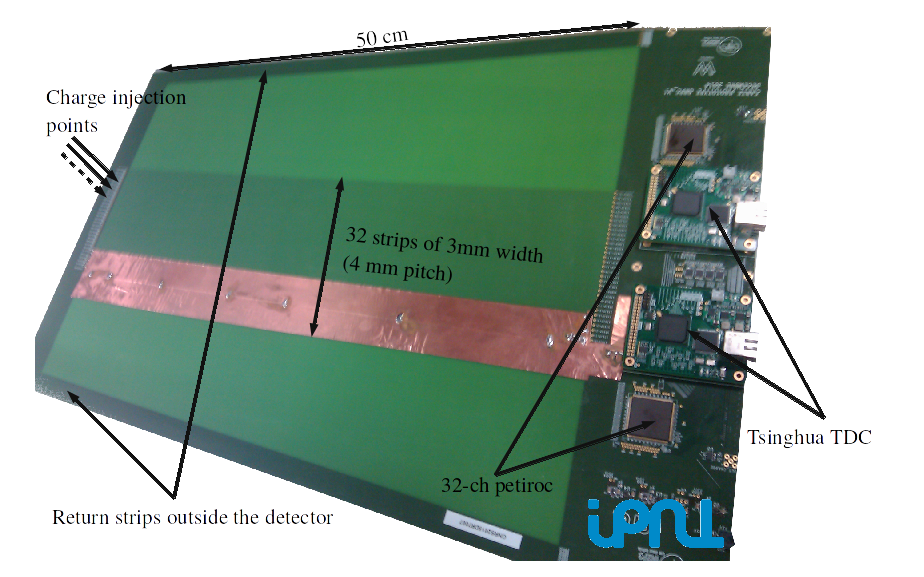}
\includegraphics[width=0.48\textwidth]{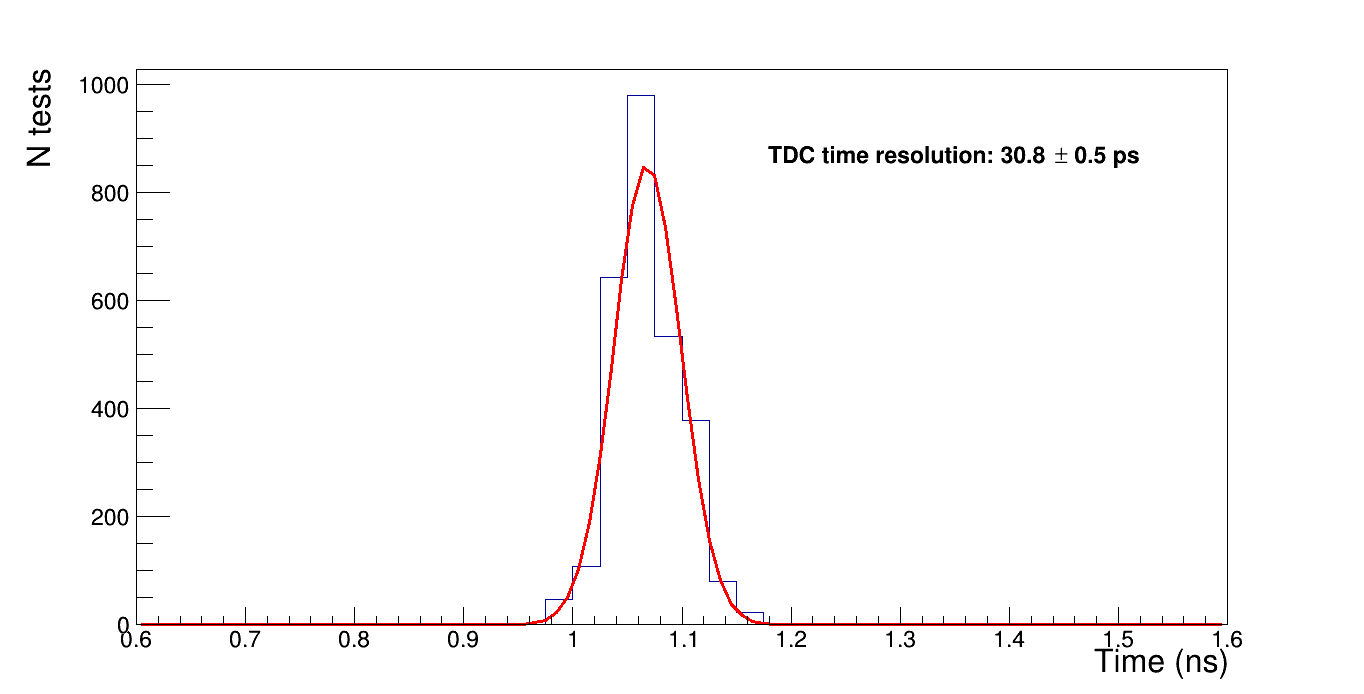}
\caption{Picture of the PCB holding PETIROC ASICs and Tsinghua TDC (top) and one of the time resolution tests using Tsinghua TDC (bottom).}
\label{PCB-photo}
\end{center}
\end{figure}

The RPC in general, and more particularly the multigap ones, are excellent fast timing detectors. To exploit this feature a dedicated electronic readout system is being developed. We propose a new kind of PCB (Figure. \ref{PCB-photo} (top)) that hosts 4.0\,mm pitch strips readout by a 32 channel ASIC called PETIROC~\cite{petiroc} with low jitter (less than 20\,ps for charges of more than 150\,fC). 

Each strips are read from both side in order to compute the position of the hit along the strip using the time arrival of the signal. A 24-channel TDC, with a time resolution of 25ns, is used to this purpose. The time resolution of the PCB is tested by injecting few thousands of times charges created by a 10V, 10ns duration square signal injected through a capacitor of 1 pF in several test points located on the strips. An example of such test is provided in Figure.~\ref{PCB-photo} (bottom). The timing distribution is fitted by a gaussian distribution. The mean of the gaussian is an irrelevant quantity for this discussion and is related to the specific properties of the electronics and the length of the connectors. The resolution of the gaussian, from 20--30\,ps, describes in contrary the intrinsic timing resolution of the PCB.

\section{Conclusion and next steps}
A new kind of GRPC detectors is proposed to equip some of the high $\eta$ muon stations of CMS. The new detector uses Tsinghua low-resistivity glass and could stand particle rate exceeding few $\khzcm$. Although the low-resistivity glass is produced in small plates, it was shown that one can build large, robust and efficient detectors. To achieve excellent time resolution measurement a new electronic board equipped with low noise ASICs and precise TDC was conceived and built. Preliminary results show that an excellent time resolution of the order of 25\,ps could be reached.  

Several steps are considered to improve our knowledge of the new detectors and finalize the design proposal. The aging properties of the Tsinghua low-resistivity glass are under study in GIF++. A monitoring of the total current is performed as function of the integrated charge. Many sessions of test beams are expected in 2016 to check the efficiency of the small detectors after an irradiation damage that will be equivalent to the one expected in HL-LHC. 
Big chambers designed in IPNL with Tsinghua glass plates mechanically fixed will also be tested in GIF++ following a similar program to the small chambers. Multigap low resistivity GRPC designed in Tsinghua will also be tested.
 Finally the PCB designed to measure the timing would be placed in between two large multigap GRPC in the future to check that a time resolution better than 50 ps could be reached.

\acknowledgments
We are grateful to the CERN EN and EP departments for the facility infrastructure support.

\end{document}